\documentclass{sig-alternate}
\usepackage{graphicx}
\usepackage{balance}  
\usepackage{booktabs}
\usepackage{xspace}
\usepackage{url}
\usepackage{footnote}
\usepackage{flushend}

\usepackage{amssymb}

\newcommand{\IC}{InstaCluster\xspace}

\makeatletter
\def\@copyrightspace{\relax}
\makeatother

\begin{document}

\title{InstaCluster: Building A Big Data Cluster in Minutes
}

\numberofauthors{2} 

\author{
\alignauthor
        Giovanni Paolo Gibilisco\\
       \affaddr{DEEP-SE group - DEIB -  Politecnico di Milano}\\
       \affaddr{via Golgi, 42}\\
       \affaddr{Milan, Italy}\\
       \email{giovannipaolo.gibilisco@polimi.it}
\alignauthor
       Sr\dj{}an Krsti\'c\\
       \affaddr{DEEP-SE group - DEIB -  Politecnico di Milano}\\
       \affaddr{via Golgi, 42}\\
       \affaddr{Milan, Italy}\\
       \email{srdan.krstic@polimi.it}
}

\maketitle

\begin{abstract}
Deploying, configuring, and managing large clusters is 
very a demanding and cumbersome task due to the complexity of such
systems and the variety of skills needed. 
One needs to perform low-level configuration of the cluster
nodes to ensure their interoperability and connectivity, as well as
install, configure and provision the needed services.

In this paper we address this problem and demonstrate how to
build a Big Data analytic platform on Amazon EC2
in a matter of minutes. Moreover, to use our tool, embedded into a
public Amazon Machine Image, the user does not need to be an expert in
system administration or Big Data service configuration.
Our tool dramatically
reduces the time needed to provision clusters, as well as the cost of
the infrastructure. Researchers enjoy an additional benefit of having
a simple way to specify the experimental environments they use, so
that their experiments can be easily reproduced by anyone using our
tool. 

\end{abstract}

\section{Introduction}
\label{sec:intro}

In the last years we have witnessed a quick growth of the market interest into Big Data related products. In particular, many open source tools, mostly coming from the Apache community, have been ported or included into different commercial distributions that ease the access to powerful analytic and computational instruments for users with little or no infrastructure management experience. To build a Big Data analytic platform from scratch, a user needs to perform four basic steps: \emph{Service Selection}, 
\emph{Cluster Provisioning}, \emph{Service Provisioning}, and \emph{Service Interaction}.

In the \emph{Service Selection} step the user gathers the requirements for the analysis that he/she wants to perform and chooses the best services among those available from the Big Data community. The result of this step is not only a set of services to be provisioned in the third step, but also a set of infrastructural requirements that drive the provisioning of the cluster.

The aim of the \emph{Cluster Provisioning} step is to provide the backbone of the entire platform, i.e., the set of resources needed to host all the services required by the analysis. The characteristics of this infrastructure highly depend on the type of analysis the user wants to perform.
Once the cluster infrastructure has been provisioned the user needs to install the selected services. We call this step \emph{Service Provisioning} and it involves manual interaction with the infrastructure in order to configure all the services and allow them to cooperate. To ease the configuration effort requiring expert knowledge of the specific services, several automated service deployment and configuration tools have been developed.

The final step to build a Big Data analytic platform is to ease the interaction between the user and the services. We call this step \emph{Service Interaction}. In order for the user to interact and have a comprehensive view on all the available services, a deep integration is needed between the user interface, typically web based, and the analytic services.

Performing the above mentioned steps poses a great challenge, because of the complexity of the infrastructure and the combination of skills needed. To select the appropriate services one needs to be proficient in data analysis. Expertise in the system administration is needed to properly configure the infrastructure. Finally, expertise and familiarity with the existing Big Data frameworks is needed to install and configure the selected services.

To the best of our knowledge no open source tools, among those presented in Section \ref{sec:stack} and many other minor contributions evaluated, provide a fully automated way to perform Cluster Provisioning and Service Provisioning allowing the user to have a fully functional Big Data analytic platform, complete with Service Interaction tools. For this reason we built \IC.

Our contribution is \emph{\IC}, a tool that provides automated support to cluster and service provisioning steps of the building process of Big Data analytic platforms. In the \emph{Cluster Provisioning} step, \IC automates the discovery and configuration of a cluster infrastructure within the Amazon cloud service.
\IC delegates service provisioning to an open source Service Provisioning tool called Ambari, that it automatically installs and configures. 
The second contribution is the integration of the popular Hue Service within Ambari. The integration of this service allows users to analyze large amounts of data and visualize the results of their analysis, without knowledge of any low-level details of Big Data services. The two contributions constitute a comprehensive and turnkey solution for building and using a Big Data analytic platform.

Note that \IC is not a traditional tool with a dedicated user interface; rather it is embedded into an Ubuntu based Amazon Machine Image that we provide to the users.

The rest of the paper is organized as follows. Section~\ref{sec:stack} introduces the Big Data stack and surveys the similar solutions. In Section~\ref{sec:approach} we describe our \IC tool. We give some concluding remarks in Section~\ref{sec:conclusion} and present our demonstration plan and some additional information about the tool in Appendix~\ref{sec:demo}.


\newpage
\section{The Big Data Stack Overview}
\label{sec:stack}


In this section we first provide a short overview of the Big Data services
that can be considered in the \emph{Service Selection} step. 
The rest of the steps can be associated with a corresponding
\emph{software system}: cluster provisioning system (CPS),
 service provisioning system (SPS) and service interaction system (SIS).

We define typical requirements of these systems and give an overview of
the existing commercial and non-commercial tools that partially satisfy
the mentioned requirements.

\subsection{Big Data Services}
\label{subsec:distributions}

The Hadoop ecosystem~\cite{hadoop} is the most popular open-source
project that offers many Big Data services. At its core it provides a
framework for distributed storage and distributed processing of very
large data sets.
The standard Apache Hadoop distribution includes:
a MapReduce framework~\cite{mapreduce} for running computations in parallel;
a Distributed File System (HDFS); and the Hadoop Common, a set of libraries and utilities
used by other Hadoop modules. The Hadoop version 2.0 also includes
YARN (Yet Another Resource Negotiator), a cluster resource
management system that negotiates and schedules resources for
multiple different distributed applications running and competing for
resources in the cluster. 


Many services are built on top of the Hadoop core
services. Table~\ref{tab:services} provides an overview of such
services and the current open-source tools\footnote{Ambari and Hue are
described in subsequent sections} that 
support their provisioning and interaction. We mark with an asterisk (*)
the fields where we provided the support as part of the contributions of this paper.
Refer to~\cite{hadoop} for an exhaustive list of Big Data services and
their detailed descriptions.

\begin{savenotes}
\begin{table}
  \centering
\scriptsize
\caption{Overview of the services supported by open-source
  provisioning and interaction tools}
\begin{tabular*} {0.48\textwidth}{@{\extracolsep{\fill} }  l c c  }
\toprule

Service &
Service provisioning &
Service interaction\\

\midrule
\emph{Apache HDFS}&
Ambari &
Hue, natively \\

\emph{Apache YARN}&
Ambari &
Hue, natively \\

\emph{Apache Tez}&
Ambari &
n/a\\

\emph{Apache Hive}&
Ambari &
Hue\\

\emph{Apache HBase}&
Ambari &
Hue \\

\emph{Apache Pig}&
Ambari &
Hue \\

\emph{Apache Sqoop}&
Ambari &
Hue \\

\emph{Apache Oozie}&
Ambari &
Hue \\

\emph{Apache Zookeeper}&
Ambari &
Hue \\

\emph{Apache Falcon}&
Ambari &
n/a\\

\emph{Apache Storm}&
Ambari &
natively\\

\emph{Apache Flume}&
Ambari &
n/a\\

\emph{Apache Slider}&
Ambari &
n/s\\

\emph{Apache Knox}&
Ambari &
n/s\\

\emph{Apache Kafka}&
Ambari &
n/s\\

\emph{Apache Spark}~\cite{spark}&
Ambari &
Hue \\

\emph{Impala}~\cite{impala}&
n/s &
Hue\\

\emph{Hue}~\cite{hue}&
n/s*&
natively\\

\emph{Nagios}~\cite{nagios}&
Ambari &
Ambari\\

\emph{Ganglia}~\cite{ganglia}&
Ambari &
Ambari, natively\\

\bottomrule

n/s - not supported &
n/a - not applicable & \\

\end{tabular*}

\label{tab:services}
\end{table}
\end{savenotes}
\normalsize

\subsection{Cluster Provisioning Systems}
\label{subsec:cps}

Cluster provisioning is the process of preparing a group of (possibly virtual)
hosts with appropriate configuration, data and software to make
them interoperable. This commonly involves setting up hostnames,
addresses, secure shell connections and installing software used for
synchronization, monitoring and debugging. Depending on particular
solutions it may also involve installing a service provisioning system
on the cluster.

Basic requirements of a cluster provisioning system (CPS) are to
support different configurations of the hosts it provisions, like CPU,
memory, storage, operating system and others. In the case of virtual
hosts, a CPS can support public, private or hybrid IaaS solutions.
CPS is also responsible for the cluster lifecycle management, i.e. it
handles the changes in the configuration of the cluster that can be
performed by adding, removing, powering up or down of the hosts.
Advanced requirements include exporting cluster configurations,
cluster cloning and configuration optimizations with respect to cost
and performance. 

To the authors' best knowledge there are three cluster provisioning
systems that satisfy some of the requirements above. 
 
\emph{The Databricks Cloud}~\cite{databricks} provides both cluster
and service provisioning capabilities for the
Apache Spark Big Data pipeline. It is hosted on Amazon AWS and gives users the
ability to start and manage clusters very quickly. However, it does not provide
support for all Big Data services. Also Databricks Cloud
offers a very limited control of the particular infrastructure used
and therefore, less opportunity for cost optimizations.
%

Cloudera Director~\cite{director} is currently the most mature available CPS. It
implements majority of the CPS requirements. It currently supports
Amazon as IaaS provider, but support for more providers is planned
for the future. However, Cludera Director is closed-source and
to be used requires a Cloudera subscription.

Another closed-source CPS solution is IBM Platform Computing~\cite{ibm}.

\subsection{Service Provisioning Systems}
\label{subsec:sps}

A service provisioning system (SPS) deals with a cluster-wide configuration,
deployment and management of multiple distributed services. The main
requirements of a SPS are installation, configuration, starting,
stopping, monitoring and removal of services on the cluster. Some SPS
also calculate the best deployment configuration based on the
selected services and cluster configuration and can export a set of
service configurations for a particular cluster.

The Apache Ambari~\cite{ambari} is an open-source service provisioning
system for Big Data
services developed by Hortonworks. It exposes a web user interface
 backed by a RESTful API that can be used for installing,
configuring, starting and stopping Big Data services across any number
of hosts in a cluster.
Ambari architecture is composed of two software components: 
a server that runs on a single machine and orchestrates the service provisioning and configuration actions;
a set of agents that run on all the machines of the cluster. 
Ambari server
monitors the cluster by receiving heartbeat messages from the
agents. It is also sends action messages to install,
configure, start or stop Hadoop services to the Ambari agents.

Other closed-source SPS for Big Data
services are Cloudera Manager~\cite{manager} and MapR Control System~\cite{mapr}.

\subsection{Service Interaction Systems}
\label{subsec:hue}

A service interaction system (SIS) is an application that provides
a user interface for invoking functionality of different
services and visualizing the obtained results. SIS can also provide 
capabilities for designing the input of a certain service.

The Hue platform~\cite{hue} is an open-source SIS that enables interaction
with a Big Data cluster. It allows one to browse 
Hadoop enabled storage, design custom queries, workflows, and jobs
to process the saved data and visualize the obtained output.
Hue consists of a Hue Server that acts as a ``container'' , hosting
all the Hue web applications. It also serves as a
communication layer between the Hue web applications and the
appropriate Big Data services in the cluster. 

Talend Open Studio~\cite{talend} is another open-source SIS that
provides an IDE for designing the service input and custom data
transformation jobs by combining pre-defined graphical components.


\section{Building a Big Data Cluster}
\label{sec:approach}
\IC is a comprehensive approach that takes into consideration the cluster and service provisioning steps discussed in Section \ref{sec:intro}. In particular, it performs automatic cluster provisioning on Amazon and performs service provisioning by means of the open-source Ambari tool which we extended to support Hue. The integration of Hue within Ambari enables the service interaction step. 

\subsubsection*{Cluster Provisioning}
After service selection, cluster provisioning is the next step of the building process of a Big Data analytic platform. Our approach targets Amazon, one of the most popular cloud service providers; in particular, it uses the IaaS resources via the Elastic Compute Cloud (EC2) web service.
The provisioning of the individual VMs is delegated to Amazon EC2, while we implement the cluster provisioning logic with a set of bash scripts\footnote{\url{http://deib-polimi.github.io/InstaCluster/}} that we embed into a dedicate Amazon machine image\footnote{Refer to the documentation on Github for the image ID}. The cluster provisioning scripts implement the basic requirements of a CPS from Section~\ref{subsec:cps} and support all the instance types (i.e., configurations) available on Amazon running Ubuntu 12.04.

\begin{figure}[ht]
    \centering
    \includegraphics[width=0.45\textwidth]{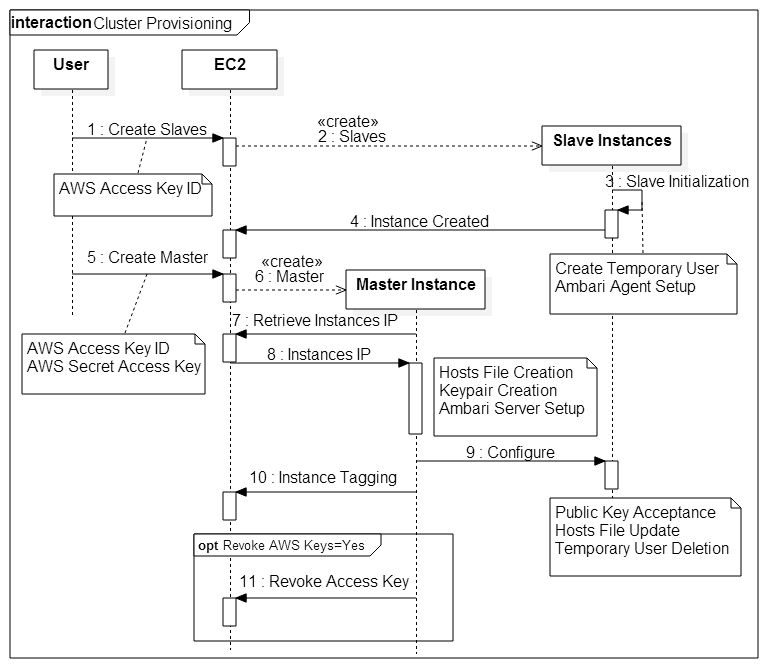}
    \caption{Cluster provisioning sequence diagram}
    \label{fig:clusterProvisioning}
\end{figure}

A provisioned cluster is composed of a number of \emph{slave} instances and a single \emph{master} instance. Slave instances are used to host Big Data services while the master instance is used to host the Ambari server and manage the cluster. The master instance can also host Big Data services, but the best practice is to have a machine dedicated only to service provisioning. 
When launching an instance, Amazon allows the user to provide some data that are accessible from the instance using its REST api. This feature can be used to trigger scripts or to provide configuration parameters to different instances generated by the same machine image. 
The main steps of the cluster provisioning are shown in Figure \ref{fig:clusterProvisioning} and explained in more details here.

The \emph{Slave} instances require the user's AWS Access Key ID passed as configuration parameter. When a \emph{slave} instance is spawned, it first creates a temporary user using the provided key ID as the password. The temporary user's credentials are then used by the master instance to distribute a newly generated key-pair that will be used during the complete lifecycle of the cluster. As a final step, the slave instances download and install the latest the Ambari agent, a piece of software needed by service provisioning tool to install and configure the required services. 

The \emph{Master} instance requires three configuration parameters. The AWS Access Key ID, that is used as password for the temporary user on slave instances, the AWS Secret Access Key, used to query EC2 for the IPs of the slave instances and the specification of the Amazon region in which to search for the slave instances. An optional fourth parameter can be specified in order to automatically make the AWS Access Key inactive as soon as the discovery of slave instances is performed. However, this is advisable only if you use spot instances, because starting and stopping instances needs a valid AWS keys. Upon booting, the master instance queries EC2 for the slave instances running in the specified region. Then it assigns a hostname to each slave instance, updates the hosts file accordingly and creates a new key-pair that will be used in the cluster. It distributes the new key-pair and the updated hosts file through secure shell using the temporary user's credentials, and triggers the deletion of the temporary user on all instances of the cluster. The master instance then informs EC2 about the naming of slaves by adding tags to each instance so that the user is able to easily identify the role of each machine in the cluster. Tagging machines in EC2 is also useful if the cluster is stopped and restarted since in such a situation the private IPs of instances might change. All the communications between the services are performed using hostnames that are assigned according to the initial enumeration performed by the master. If a restart of the cluster is performed and new IPs are assigned, the master is able to query EC2, if AWS Access and Secret Keys are still active, update the new private IPs of the slave instances in the hosts file and redistribute the new hosts file to the slave instances again. As final step the master installs and starts the Ambari server.

From the security standpoint the default Ubuntu security rules are applied during the whole cluster provisioning period. The temporary user and password-based authentication is allowed until the master instance discovers the slave instances and shares a generated key-pair. As soon as the key-pair is shared between the instances, the default authentication method is restored and the temporary user is deleted. The additional key-pair introduced to authenticate the Ubuntu user is unique for each cluster spawn by the image and is revoked and regenerated after each restart of the entire cluster. Besides using the generated key-pair, all the instances can be accessed with the user's own key-pair downloaded from Amazon.

\subsubsection*{Service Provisioning}
The \emph{service provisioning} step is delegated to Ambari. The configuration of the cluster performed automatically in the first step allows the tool to discover and interact with slaves with minimum user intervention. In this step the user has to provide its own key, generated by Amazon and specify the username to use when connecting to slaves. Ambari is installed automatically and started on port 8080 of the master instance. After the slave discovery performed by the Ambari server, which identifies the agents installed in the previous step, the user needs to select the services he/she selected, check or modify the suggested configuration and let the tool set them up. 

The initialization of master and slave nodes performed by the provided image in the previous step ensure that Ambari communications, performed using hostnames and passwordless ssh with key-pairs, do not encounter any problems. We have also extended the installation of Ambari to offer provisioning of the standalone Spark and Hue services.  

The choice of configuration parameters for the services to be installed is suggested to the user by Ambari and can be changed by the user if needed. 
The configuration ports for the additional services integrated in this work are listed in Table \ref{tab:ports}.

\begin{table}
 \centering
\scriptsize
\caption{Ports of the additional services}
\begin{tabular*} {0.48\textwidth}{@{\extracolsep{\fill} }  l c   }
\toprule

Service Name 		&  Port Number \\ 

\midrule 

Spark Driver 		&  7077 \\
Spark Web UI 		&  8888 \\
Spark Job Server 	&  8090 \\
Hue Web UI			&  8808 \\ 
\bottomrule

\end{tabular*}

\label{tab:ports}
\end{table}
\normalsize

\subsubsection*{Service Interaction}
The support for the interaction with the installed services 
is provided by Hue. We chose Hue due to our extensive familiarity 
with its functionality. If the user chooses to install Hue through Ambari, we make sure that the configuration of Hue 
correctly targets each service installed by Ambari and run it on port 8808.


\section{Conclusion and Future work}
\label{sec:conclusion}

This demonstration presents \IC, an open-source tool that automates cluster
provisioning and service provisioning steps of the building process of Big Data analytic platforms. 
Using \IC, researchers can produce repeatable experiments by sharing with the community their code, the
input data, the size of the cluster (in terms of type and number of
VMs) and any configuration of the parameters that is changed with
respect to the default ones. 
The provided solution allows the use of spot instances in order to
further reduce experimental costs.
The main advantage of using \IC is the reduction in time and expertise
needed to setup a cluster allowing researchers and users in general to
focus on more productive aspects.
A key difference between \IC and similar tools is the fact that, at the best of our knowledge, \IC is the only completely open-source tool to automatize all the steps needed to provision a complete Big-Data analytic platform.
Using \IC we have managed to build a small size cluster,
composed by 4 VMs of type \emph{c4.xlarge} hosting all the supported
services from Table \ref{tab:services} in 25 minutes. An experienced
system administrator, without the help of \IC  would need several
hours to build an equivalent cluster and the process would be highly
involving and error-prone.  

The main limitation of the tool is that it currently supports one cluster
per Amazon region, we plan to extend the support for multiple
clusters per region. Future directions involve supporting other operating systems
and other IaaS providers, as well as providing a fully automatic way
to configure services without interacting manually with Ambari.


\section{Acknowledgments}
Authors acknowledge the MODAClouds European project supported the European Commision grant no. FP7-ICT-2011-8-318484 for supporting this project with the use of Amazon EC2 resources. Authors would also like to acknowledge the Ambari user list for the great support received. 
\bibliographystyle{abbrv}
\bibliography{insta}  

\pagebreak

 \begin{appendix}
\section{Demonstration Plan}
\label{sec:demo}

 
Our demo will cover eight use cases that our two contributions
enable. The goal of these use cases is to show how a researcher or
practitioner can simplify the experimental setup and what data he/she
needs to report in order for anyone to reproduce the experimental
results. A video showing the usage of the tool is also available on the tool repository. 

\textbf{Use case 1:} 
The user selects services that he/she intends to use (e.g. Spark and MapReduce), provisions a
6 node cluster and installs all the selected services.
  
\textbf{Use case 2:} 
Once all the services are installed 
the user stops all the instances of the cluster to prevent unnecessary billing.

\textbf{Use case 3:} 
User starts the cluster. When starting the instances it is important to
start slave instances first.

\textbf{Use case 4:} 
The user extends the cluster by adding three more machines. The cluster
is stopped and user first creates the three additional instances,
starts the rest of the slave instances and then starts the master.

\textbf{Use case 5:}
The user uses Hue to browse Hadoop enabled storage (e.g. HDFS).

\textbf{Use case 6:}
The user uses Hue to submit a Spark job.

\textbf{Use case 7:}
The user uses Hue to upload a file to HDFS.

\textbf{Use case 8:}
The user uses Hue to execute a MapReduce WordCount job over the file
uploaded to HDFS.

\section{Tool information}
\label{sec:tool}
The tool is available on
github\footnote{https://github.com/deib-polimi/InstaCluster} as a set
of open source Shell and Python scripts that can be installed on any
linux machine running on AWS. For great convenience of users an Amazon
Machine Image with pre-installed scripts is available\footnote{Refer
  to the github documentation for the most updated AMI id}. 
The documentation on the usage of the tool, either in the script
format or in the AMI format, can be found on
github\footnote{http://deib-polimi.github.io/InstaCluster/}. A quick
video that guides the user through the interaction with the tool is
available\footnote{https://www.youtube.com/watch?v=Vqu0cjQ7M0w}  
The tool has been used by several Computer Science PhD and Master
students at Politecnico di Milano in order to perform repeatable
experiments on a cloud environment in a cost effective manner.


 \end{appendix}

\end{document}